\documentclass[default,iicol,sn-mathphys]{sn-jnl}

\jyear{2021}%

\usepackage{xr}
\begin{document}

\title[a]{Sequential time-window learning with approximate Bayesian computation: an application to epidemic forecasting}

\author*[1]{\fnm{João Pedro} \sur{Valeriano}}\email{joaopedrovm98@gmail.com}

\author[2]{\fnm{Pedro Henrique} \sur{Cintra}}\email{pedrohpc@ifi.unicamp.com}

\author[3]{\fnm{Gustavo} \sur{Libotte}}\email{glibotte@lncc.br}

\author[4]{\fnm{Igor} \sur{Reis}}\email{igorreis.0203@gmail.com}

\author[5]{\fnm{Felipe} \sur{Fontinele}}\email{feradofogo@gmail.com}

\author[3]{\fnm{Renato} \sur{Silva}}\email{rssr@lncc.br}

\author[3]{\fnm{Sandra} \sur{Malta}}\email{smcm@lncc.br}

\affil*[1]{\orgdiv{Instituto de Física Teórica}, \orgname{Universidade Estadual Paulista}, \orgaddress{\street{R. Dr. Bento Teobaldo Ferraz, 271, Bloco 2, Barra Funda}, \city{São Paulo}, \postcode{01140-070}, \state{São Paulo (SP)}, \country{Brazil}}}

\affil[2]{\orgdiv{Instituto de Física "Gleb Wataghin"}, \orgname{Universidade Estadual de Campinas}, \orgaddress{\street{Rua Sérgio Buarque de Holanda, 777}, \city{Campinas}, \postcode{13083-859}, \state{São Paulo (SP)}, \country{Brazil}}}

\affil[3]{\orgname{Laboratório Nacional de Computação Científica}, \orgaddress{\street{Av. Getulio Vargas, 333}, \city{Petrópolis}, \postcode{25651-076}, \state{Rio de Janeiro (RJ)}, \country{Brazil}}}

\affil[4]{\orgdiv{Instituto de Física}, \orgname{Universidade de São Paulo}, \orgaddress{\street{Av. Trab. São Carlense, 400 - Parque Arnold Schimidt}, \city{São Carlos}, \postcode{13566-590}, \state{São Paulo (SP)}, \country{Brazil}}}

\affil[5]{\orgdiv{Department of Physics}, \orgname{University of Alberta}, \orgaddress{\street{116 St \& 85 Ave}, \city{Edmonton}, \postcode{T6G 2E1}, \state{Alberta}, \country{Canada}}}


\abstract{The long duration of the COVID-19 pandemic allowed for multiple bursts in the infection and death rates, the so-called epidemic waves. This complex behavior is no longer tractable by simple compartmental model and requires more sophisticated mathematical techniques for analyzing epidemic data and generating reliable forecasts. In this work, we propose a framework for analyzing complex dynamical systems by dividing the data in consecutive time-windows to be separately analyzed. We fit parameters for each time-window through an Approximate Bayesian Computation (ABC) algorithm, and the posterior distribution of parameters obtained for one window is used as the prior distribution for the next window. This Bayesian learning approach is tested with data on COVID-19 cases in multiple countries and is shown to improve ABC performance and to produce good short-term forecasting.}

\keywords{Covid-19, Epidemic forecasting, Approximate Bayesian Computation, SEIRD model}



\maketitle

\section{Introduction}
\label{intro}

Since the onset of the novel Coronavirus (SARS-CoV-2) pandemic, computational methodologies have played a fundamental role in helping to understand the dynamics of the spread of the virus in society \cite{sonabend2021non}. Computational models are capable of capturing, to a certain extent, the behavior of the data that describes the advance of the virus, making it possible to simulate predictive scenarios that collaborate with the decision-making of government authorities and in the allocation of medical and financial resources. Mathematical models and computer simulations can also provide relevant indicators to assist in the implementation of social distancing measures, hoping to stave off the advance of the disease. Adiga~et~al.~\cite{bib:adiga2020} present a comparative analysis of computational models used to describe the behavior of the epidemic. Challenges of modeling COVID-19 are discussed in Refs.~\cite{bib:bertozzi2020,bib:vespignani2020}, whereas Eker~\cite{bib:eker2020} analyzes the validity and usefulness of computational models in such context.

To date, the world has had more than 378 million confirmed cases, with more than 5{,}69 million individuals dead~\cite{bib:owidcoronavirus}. In several countries, the number of daily cases has already had at least two waves of infection, when a meaningful increase in the number of cases occurs after a significant drop in the number of new infections during the previous wave. Numerous compartmental models, which have been widely used to simulate the spreading of COVID-19~\cite{bib:massonis2020}, in its canonical form, have no descriptive capacity to represent the behavior of data with multiple waves~\cite{bib:moein2021,bib:brauer2008}. Further drawbacks of the classical SIR model are discussed by Singh and Gupta~\cite{bib:singh2021}. This poses an even greater challenge when such models are used to simulate the spreading dynamics of COVID-19, requiring more sophisticated computational frameworks to be established, to provide more reliable results.

A growing body of literature has proposed computational models and techniques to overcome the difficulties imposed by the data when the epidemic is at an advanced stage. Of note, the variety of works related to the modeling of the second (and subsequent) waves of COVID-19 is more restricted than those related to the early stages of the pandemic. Below, we summarize the most relevant ones that we are aware of. Kaxiras and Neofotistos~\cite{bib:kaxiras2020a} extended the forced-SIR model, proposed in Ref.~\cite{bib:kaxiras2020b}, which provides an approximate analytical solution for the differential equations that represent the well-known SIR model, to allow multiple waves to be captured; Cacciapaglia~et~al.~\cite{bib:cacciapaglia2021} models the multi-wave pattern by considering a master equation for the time-evolution of the total number of infected individuals in particular locations. Such equation is based on the epidemic Renormalization Group (eRG) framework~\cite{bib:cacciapaglia2020}, which is extended to include the diffusion of the epidemic between multiple nearly-isolated regions; Singh and Gupta~\cite{bib:singh2021} propose what they call the Generalized SIR (GSIR) model, which is an integrative model encompassing multiple waves that emerge and vanish within a time interval. Special solutions of the constituent waves of the model are demonstrated, employing well-known growth functions, leading to time-varying parameters and a closed-form solution of all the system parameters.

\section{Motivation and objectives}

As mentioned before, although there is no current closed definition for an infection wave, several countries have had more than one sequence of strong increase followed by a substantial drop in the number of daily new cases, which is popularly characterized as an infection wave. Typical compartmental models (such as SIRD and SEIRD) are not capable of capturing this behavior considering its canonical structure \cite{bib:moein2021inefficiency}. Such restriction is a result of the basic assumptions behind the model, that the population is homogeneously mixed, resulting in one single infection wave until the so-called ``herd immunity'' is reached.

To account for inhomogeneous mixing in the population, reinfection due to poor immune response or immunity loss, specific social behaviors, or governmental policies that can change the infection dynamics, several groups work with modified SIRD/SEIRD models \cite{batistela2021sirsi,ramezani2021novel,asamoah2021sensitivity}. However, adding more compartments may drastically increase the number of parameters to be fitted in the model. For instance, Ramezani~et~al.\cite{ramezani2021novel} implements a modified SEIRD model to account for asymptomatic patients and individuals who self isolate (SEARIDQ model), which uses a total of 14 parameters. Such an increase in the number of parameters also increases the chances of falling into a non-identifiable model, using the same dataset~\cite{bib:massonis2020structural}, given the same number of curves to be fitted. Although some techniques have been proposed to bypass this problem, they often require more data than what is available.

Another approach for capturing the complex dynamics of local epidemics is to use SIRD/SEIRD models with time-varying parameters. For example, if $\beta$ corresponds to the infection rate of susceptible individuals, the use of masks or social isolation, therefore, decreases its value \cite{he2020seir,calafiore2020time}. Despite that, the introduction of time-varying parameters usually requires confining their variation to an analytic function, which may not represent the true temporal dynamics of parameters, once many of them are not directly measurable. This choice also affects the generality of the model, as a particular choice of the functional form that describes a parameter may not apply in another context. Furthermore, compartmental models struggle to take into account testing and contact tracing in their dynamics, which further complicates the use of time-varying functions \cite{bib:sturniolo2021testing}. Even if we overcome this problem, the individual policies, social behavior and testing of each country should make the generalization of these models for other countries nearly impossible.

The challenge faced by epidemic models also increases, as new variants with higher transmissibility or immune escape emerge, such as the variants of concern (VOC) Alpha (B.1.1.7), Beta (B.1.351), Delta (B.1.617.2), Gamma (P.1), and Omicron (B.1.1.529). The appearance of each VOC is associated with local or global change in the temporal dynamics of parameters associated with the pandemic \cite{davies2021estimated, naveca2021covid}. For example, the Alpha variant is associated with a 50\% increase in transmissibility. Such an increase may reflect on a change in $\beta$ over time as the variant spread through a region~\cite{Lee2021SARSCOV2}.

Aiming to provide an alternative to fitting the limited amount of data and producing accurate short-term predictions, we propose a time-window SEIRD model, with time-varying window size as the rate of effective parameter change is not the same throughout the epidemic, and different window sizes may be more appropriate at different times. The parameters of the model may be considered constant through the time-window being fitted (see Methodology \ref{sec:time_window_method} for further details) and the number of parameters of the model remains the same, decreasing the chances of falling into a non-identifiable problem. This procedure allows capturing the temporal variations of epidemiological parameters along time-windows without requiring the model to be defined with time-dependent parameters, making it possible to fit a curve with a simple model with piece-wise constant parameters within each time-window, emulating the behavior of time-varying parameters, but not defined by an analytic function. Time-window methods are common in nowcasting (correction for reporting delays) methods for epidemiological surveillance \cite{rotejanaprasert2020bayesian,mcgough2020nowcasting,bastos2019modelling}.

To the best of our knowledge, the framework proposed by Liao~et~al.~\cite{bib:liao2020} is the one that most resembles what is being proposed here. Although both methodologies use an approach in which data is divided into time-windows, the fundamental difference is that the methodology of Liao~et~al.~\cite{bib:liao2020} uses an exhaustive search associated with the least-squares method to determine the optimal parameters of the compartmental model, and a machine learning method is employed to track and predict the values of parameters, based on the variation of the values of the basic reproduction number and a growth rate in the historical data. On the other hand, we adopt a Bayesian approach, so that the knowledge obtained from past windows is propagated to the later windows, to gradually fit the data and compose the behavior of the model.

Keeping in mind the choice of using time-windows to analyze data, the first idea might be to deal with each window separately and fit every one of them independently. As we will show, this can be an inefficient approach, and we propose an alternative solution to connect information between consecutive time-windows and use this to improve model fitting. For this purpose, we need an inference algorithm capable of using information acquired in a previous window to fit the next one. 


In this work, we choose to use ABC-SMC (Approximate Bayesian Computation with Sequential Monte Carlo) \cite{bib:minter2019approximate}, which generates a posterior distribution for the parameters of the model in an arbitrary window. This posterior distribution can then be used as the prior distribution for the next time-window, and this procedure goes for every following time-window in the dataset.

In the Results section, we present the fitting of data on COVID-19 cumulative cases and deaths in Brazil, as a proof of concept of the improvement gained by fitting time-windows using past window posteriors instead of flat priors.

\section{Methodology}

\subsection{Time-window fitting and the use of past window posteriors}
\label{sec:time_window_method}

Our goal is to analyze an epidemiological time series of cumulative infected and dead individuals, considering a model of coupled ordinary differential equations. We consider long enough time periods over which the data spam over, such that the epidemiological parameters change over time. Such a change can be due to a particular social behavior, governmental policies, environmental factors, or natural selection---all of which may lead to the emergence of multiple epidemic waves. In this case, one may suppose that the principles for the system's time evolution are the same, but some of its properties have changed over time, that is, the model is the same over the time series, although the parameters probably change.

As we are not interested in a functional form for the time variation of the model parameters, we take an alternative approach. If one considers only a small enough time interval of the dataset, then this interval should be reasonably described by a model with constant parameters. Motivated by this fact, we divide the epidemic data into multiple time-windows, each to be fitted separately with the same model, but obtaining different sets of piece-wise constant parameters.

The fitting algorithm starts by choosing a number $N$ of days to be considered in each time-window. We also need to choose by how many days one window shall be shifted from the previous. This shift will be denoted by $d$ (days). Therefore, if the first time-window goes from day 1 until day $N$, the second time-window will go from day $1+d$ until day $N+d$. Notice that, if $d < N$, there will be an overlap of $N-d$ days between consecutive windows. We fit the model to the data of a time-window using the ABC-SMC algorithm, that generates a posterior distribution for the parameters, which in turn can can be used to make predictions for periods following the end of the current time-window.


The use of the ABC-SMC algorithm for fitting the model implies the choice of a prior distribution for the model's parameters. For simplicity, for each time-window, one can start by adopting an uniform prior distributions for all parameters (with different ranges depending on the nature of each parameter), in the case of lack of knowledge to build more informative priors.

We propose a way to go beyond the flat prior simplification, still without considering the knowledge gained from the data, but only the knowledge obtained while fitting data on past time-windows. By hypothesis, if we consider that the data is described by an ordinary differential equation model with time varying parameters, the difference between the distributions of such parameters for consecutive time-windows should be small, especially in the case that an overlap exists between consecutive time-windows. Therefore the posterior distribution obtained for the $n$-th window should provide a reasonable initial estimate---the prior distribution---for the $(n+1)$-th window. So we propose to use this approach instead of a flat prior in order to provide useful information for the ABC-SMC fitting algorithm, further optimizing the process.

\subsection{Adaptive window size}

A possible problem with dividing the epidemic data into multiple time-windows is how to choose the window sizes. It is important to notice that varying window sizes may be more appropriate for different windows of the time series. To counter that, we developed a simple algorithm to choose the window size of the $n$-th window based on the goodness-of-fit in the past two windows.

First, a given size $s_{1}$ is chosen for the first time-window, and we set the lower and upper bounds for window sizes, denoted by $s_{\mathrm{min}}$ and $s_{\mathrm{max}}$, respectively. In turn, the step size in the window size variation, $\Delta s$, is chosen to be the same as the offset $d$ between the last days of consecutive time-windows. Then, the second time-window will have the same size as the first one, that is, $s_{2}=s_{1}$. But, starting from the third time-window, the window size will be chosen by the following algorithm: let $y_{m}^{i}$ be the actual data for the $m$-th day of the time-window, whereas $\hat{y}_{m}^{i}$ denotes the prediction of the model for the same day. If $n$ indexes the size $s_{n}$ of the $n$-th time-window, the Normalized Root Mean Square Deviation (NRMSD) for the $i$-th component of the data vector---denoted by $\varepsilon^{i}_{n}$---over the same window is given by

\begin{align}
\varepsilon^{i}_{n} = \frac{\displaystyle\sqrt{\sum_{m} \frac{(\hat{y}^{i}_{m} - y^{i}_{m})^2}{s_{n}}}}{y^{i}_{\mathrm{max}}-y^{i}_{\mathrm{min}}} \; ,
\end{align}

\noindent so that

\begin{align}
\varepsilon_{n} = \sum\limits_{i} \varepsilon^{i}_{n} \; ,
\end{align}

\noindent where $i$ identifies the component of which the NRMSD is being calculated, the index $m$ runs over the days inside the $n$-th time-window. Then, for the $n$-th window, with $n \geq 3$, the window size is chosen according to Algorithm \ref{alg:window_size_update}.

\begin{algorithm}
\caption{Window size selection for the $ n $-th time-window, for $ n \geq 3 $.}
\label{alg:window_size_update}
\begin{algorithmic}
\If{$\varepsilon_{n-1} \leq \varepsilon_{n-2}$}
    \If{$s_{n-1} < s_{\mathrm{max}}$}
        \State $s_{n} \gets s_{n-1}+\Delta s$
    \Else
        \State $s_{n} \gets s_{\mathrm{max}}$
    \EndIf
\Else
    \If{$s_{n-1} > s_{\mathrm{min}}$}
        \State $s_{n} \gets s_{n-1}-\Delta s$
    \Else
        \State $s_{n} \gets s_{\mathrm{min}}$
    \EndIf
\EndIf 
\end{algorithmic}
\end{algorithm}

The intuition behind this procedure is that smaller time-windows are easier to fit. In this case, we measure the goodness-of-fit by $\varepsilon$, so that the smaller $\varepsilon$, the better the quality of the fit. Therefore, if $\varepsilon$ increases from one time-window to the next, it can be understood that the fitting may require a greater deal of effort. If we assume that our model should give a good description of the data in a small enough time span, we could expect to improve the quality of the fit by making the time-window smaller, the way we proceed to the next time-window. On the other hand, if $\varepsilon$ decreases between two time-windows, recalling that consecutive windows, $n$ and $(n+1)$, have an overlap of $s_{n+1} - \Delta s$ points, we can understand that the new $\Delta s$ points at the end of the $(n+1)$-th window are in good agreement, in terms of model parameters, with the behavior of the data in the $n$-th window. Therefore, increasing the window size can allow the simultaneous consideration of a larger range of the time-series that is related to the same set of parameters of the chosen model, decreasing the possibility of overfitting and improving generalization.

The lower bound $s_{\mathrm{min}}$ can be set considering the number of free parameters in the fitting problem, bearing in mind that fitting very few data points can lead to overfitting, so it is reasonable to have at least more data points than free parameters. For the upper bound $s_{\mathrm{max}}$, it is more complex to set a strict natural limit, but it is worth recalling the motivation regarding the approach to divide the data into time-windows: there is a limit on how long a fixed set of parameters can adequately fit the data, so we set an upper bound on the maximum expected range describable by a single constant set of parameters.

Section S2 of the Supplementary information text presents a practical comparison between considering fixed and adaptive window sizes, showing that the results are rather similar, but the adaptive window size Algorithm \ref{alg:window_size_update} does not require one to choose a specific window size.

\begin{figure*}[hbtp]
\centering
\includegraphics[width=\linewidth]{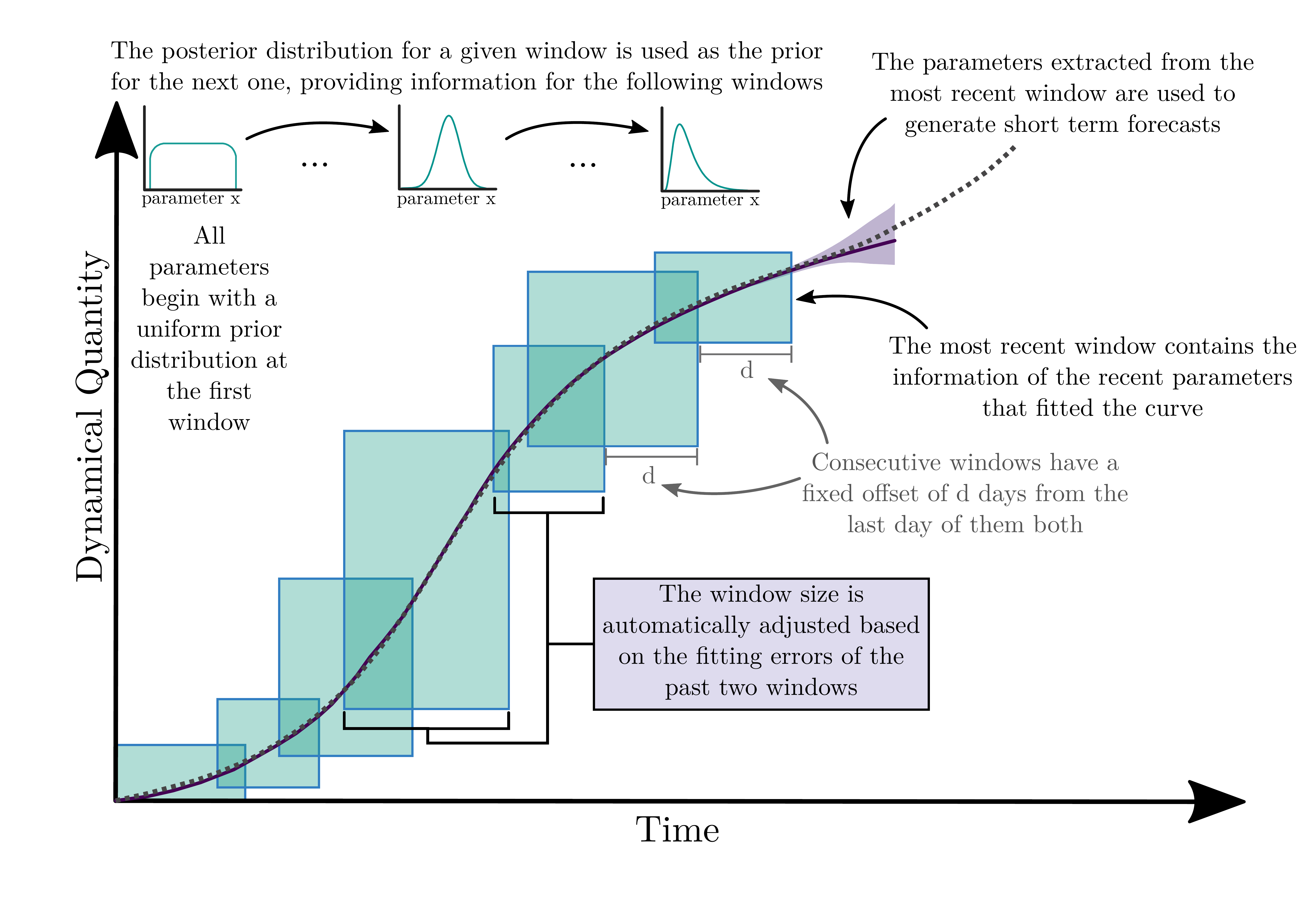}
\caption{Visualization of the framework. The data (dotted curve) of a dynamical variable, such as the cumulative reported cases of an epidemic, is fitted using windows of adaptive size (blue boxes). The prior of the parameters of the model are given by flat distributions in the first window, and the obtained posterior distribution is used as the prior to the following window. At the end, the past window generate predictions (dark purple curve) using the best parameters retrieved from the fit of the past window. From these predictions, a variability region is constructed (shaded purple region).}
\label{fig:scheme}
\end{figure*}

Figure \ref{fig:scheme} graphically summarizes the methodology described in this section for the inference of model parameters and generation of forecasts in each time-window of the the considered data.

\section{Application to Epidemic Forecasting}

We implemented the time-window model with an ABC-SMC algorithm for curve fitting. This means that we divide the epidemic curve into multiple time-windows, which are considered separately by fitting a time-independent compartmental model. The epidemic model chosen is a SEIRD model including infection by pre-symptomatic individuals (for details see \cite{loli2020monitoring,rapolu2020time}) described by the ordinary differential equations system \eqref{eq:seird}.

$\beta_I$ and $\beta_E$ stand for the infection rate of infected and exposed individuals, respectively, $c$ represents the inverse of the incubation period, $\gamma$ and $\mu$ express the recovery and death rates, respectively. The model is solved using a 4th order Runge-Kutta algorithm subjected to the constraint $N = S +E + I + R + D$, and with the initial conditions $S(0) = S_0 $, $E(0) = E_0 $, $ I(0) = I_0$, $R(0) = R_0 $, and $D(0) = D_0 $. All five parameters are set free for the fitting process, alongside the total population $N$, from which the initial condition for $S$ is derived according to $S_0 = N - I_0 - E_0 - R_0 - D_0$.

\begin{align}
\label{eq:seird}
\begin{aligned}
& \frac{\mathrm{d}S}{\mathrm{d}t} = -\frac{\beta_I S I}{N} - \frac{\beta_E S E}{N} \\
& \frac{\mathrm{d}E}{\mathrm{d}t} = \frac{\beta_I S I}{N} + \frac{\beta_E S E}{N} - \alpha E \\
& \frac{\mathrm{d}I}{\mathrm{d}t} = \alpha E - (\gamma + \mu)I \\
& \frac{\mathrm{d}R}{\mathrm{d}t} = \gamma I \\
& \frac{\mathrm{d}D}{\mathrm{d}t} = \mu I \; .
\end{aligned}
\end{align}



For our analysis, we consider data on cumulative cases and deaths for different countries. Therefore, at the beginning of each time-window, we only have initial values for deaths $D_{0}$ and cumulative cases $C_{0}$. We need a way to estimate the initial values $E_{0}$, $I_{0}$ and $R_{0}$. For doing so, we define new parameters $c_{E}$ and $c_{R}$ to be fit together with the system of equations~\eqref{eq:seird}, such that $R_{0} = c_{R} (C_{0}-D_{0}) \Rightarrow I_{0} = (1-c_{R}) (C_{0}-D_{0})$, and $E_{0} = c_{E} (C_{0}-D_{0})$.

\section{Results and Discussion}
\label{sec:results}

\begin{figure*}[hbtp]
\centering
\includegraphics[width=\linewidth]{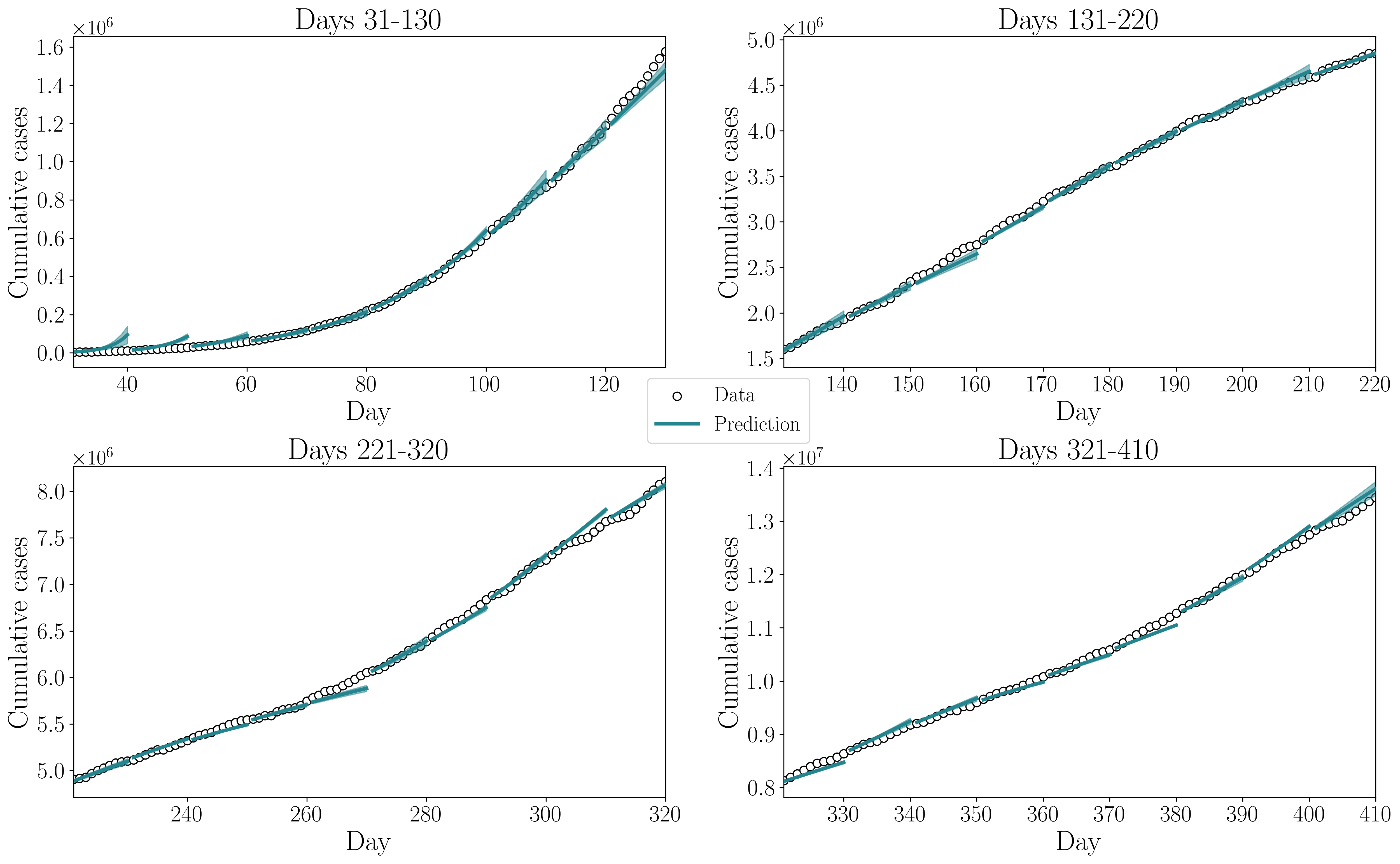}
\caption{Prediction for 10 days into the future along with the epidemic data for comparison, obtained via the past window approach. The results were split in four subplots for clarity of details. Small margins show fluctuations over the best results obtained by 10 different executions.}
\label{fig:brazil_30_10day_pred_past}
\end{figure*}

Here, we present different comparisons between results from flat prior and past window posterior approaches, fitting the SEIRD model to epidemic data on cumulative cases and deaths of COVID-19 in Brazil. To run the ABC-SMC with adaptive time-window sizes, we set the minimum time-window length $s_{\mathrm{min}} = 10$ days and the maximum window length to $s_{\mathrm{max}} = 50$ days. 

Before proceeding to the comparison between different approaches, we can already see, in Figure \ref{fig:brazil_30_10day_pred_past}, the piece-wise 10-day predictions from fitting throughout the curve of  cumulative cases of COVID-19 in Brazil. The curve is divided into four subplots for better visualization. Since windows are shifted by five days and predictions are computed for ten days, we only show forecasts of alternated windows, in order to avoid overlap in prediction curves. Although the first predictions tend to overestimate the growth due to a lack of information regarding the epidemic parameters, the remaining predictions describe the epidemic curve fairly well, capturing the general trend of cases over different scenarios. In supplementary figures S5 and S6 one can also see the fit, and the following forecast, for each separate time-window along the epidemic curve of Brazil for both approaches. In these windows, it is possible to see that the past window posterior approach leads to more consistent fittings, with smaller variation between different runs of the method.

\begin{figure}[htpb]
\centering
\includegraphics[width=\linewidth]{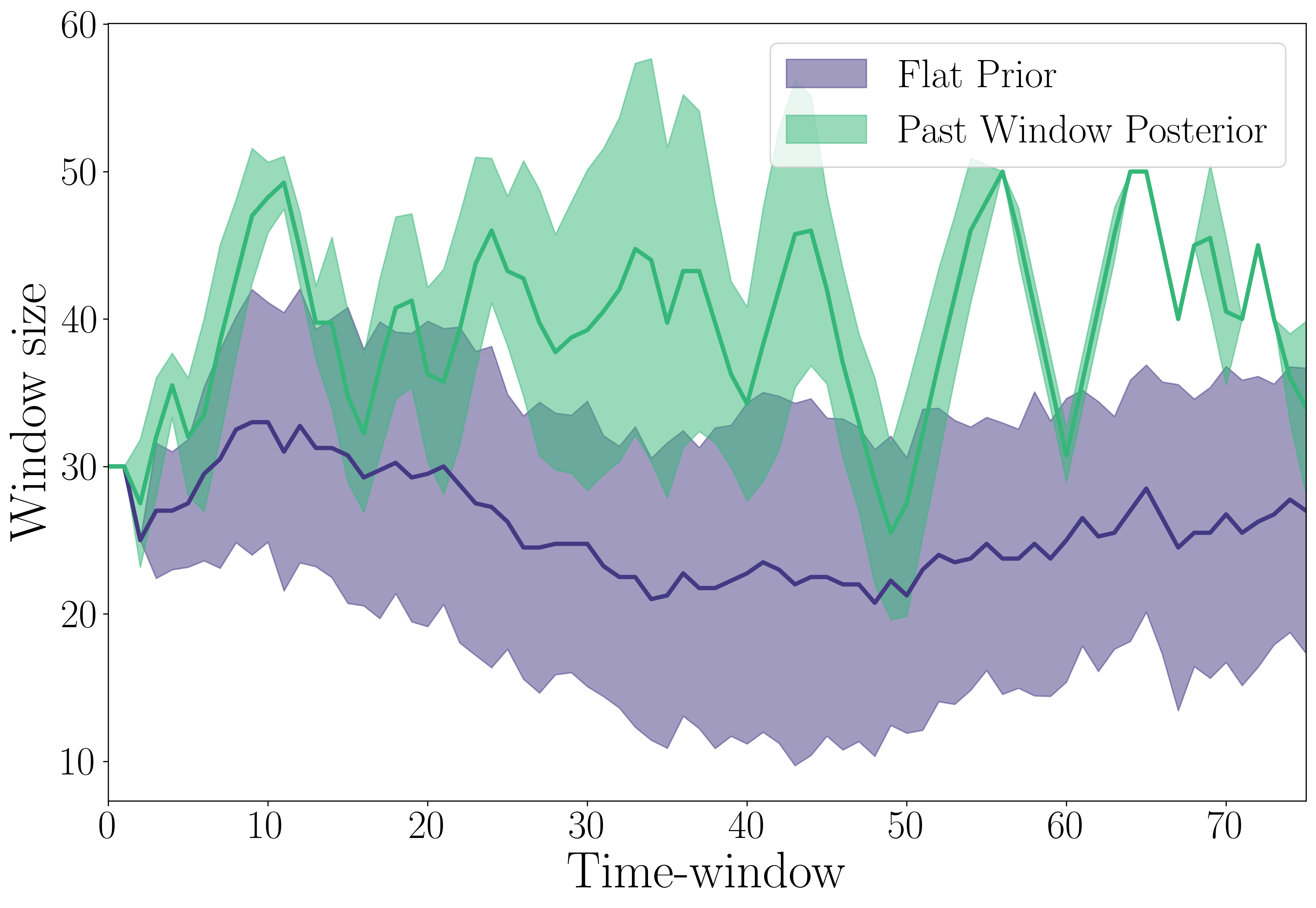}
\caption{Windows' sizes selected over five executions, for the case past window posteriors.}
\label{fig:brazil_30_window_size}
\end{figure}

Figure \ref{fig:brazil_30_window_size} shows the mean, with standard deviation, of the windows' sizes, by window, over 10 executions of the ABC-SMC fitting, starting from a 30-day time-window. More important than the actual average window size, is the fluctuation around it. The window size selection algorithm presents a better convergence to the optimal window size when combined with the past window posterior approach. This can also be seen as a hint to the convergence improvement of the ABC-SMC by the use of past window's posteriors instead of flat priors.

\begin{figure}[hbtp]
\centering
\includegraphics[width=\linewidth]{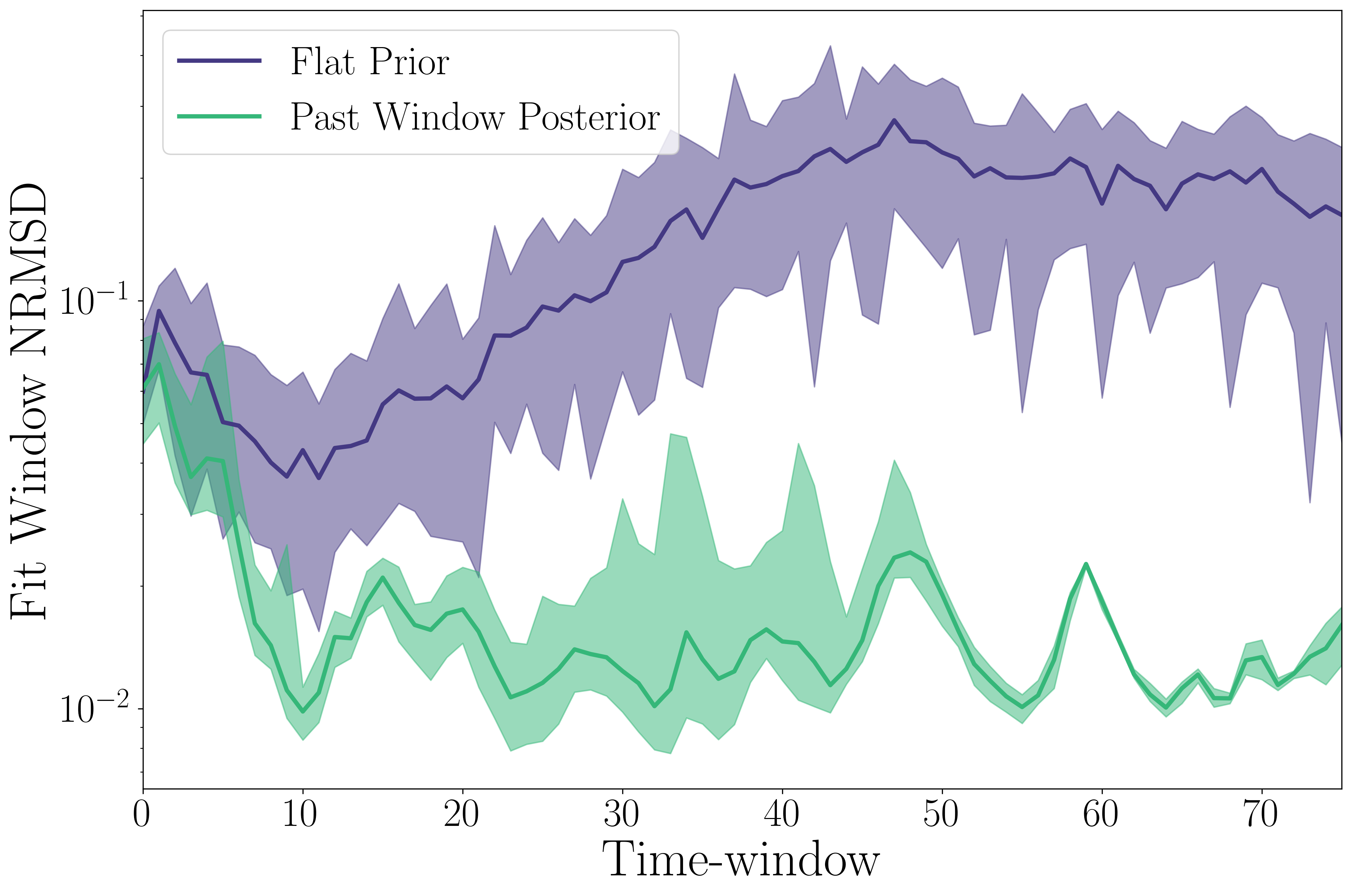}
\caption{Normalized RMSD over the fitting window, for both cases of using flat priors or past window posteriors, considering five executions for each case.}
\label{fig:brazil_30_fit_window_nrmsd}
\end{figure}

\begin{figure}[hbtp]
\centering
\includegraphics[width=\linewidth]{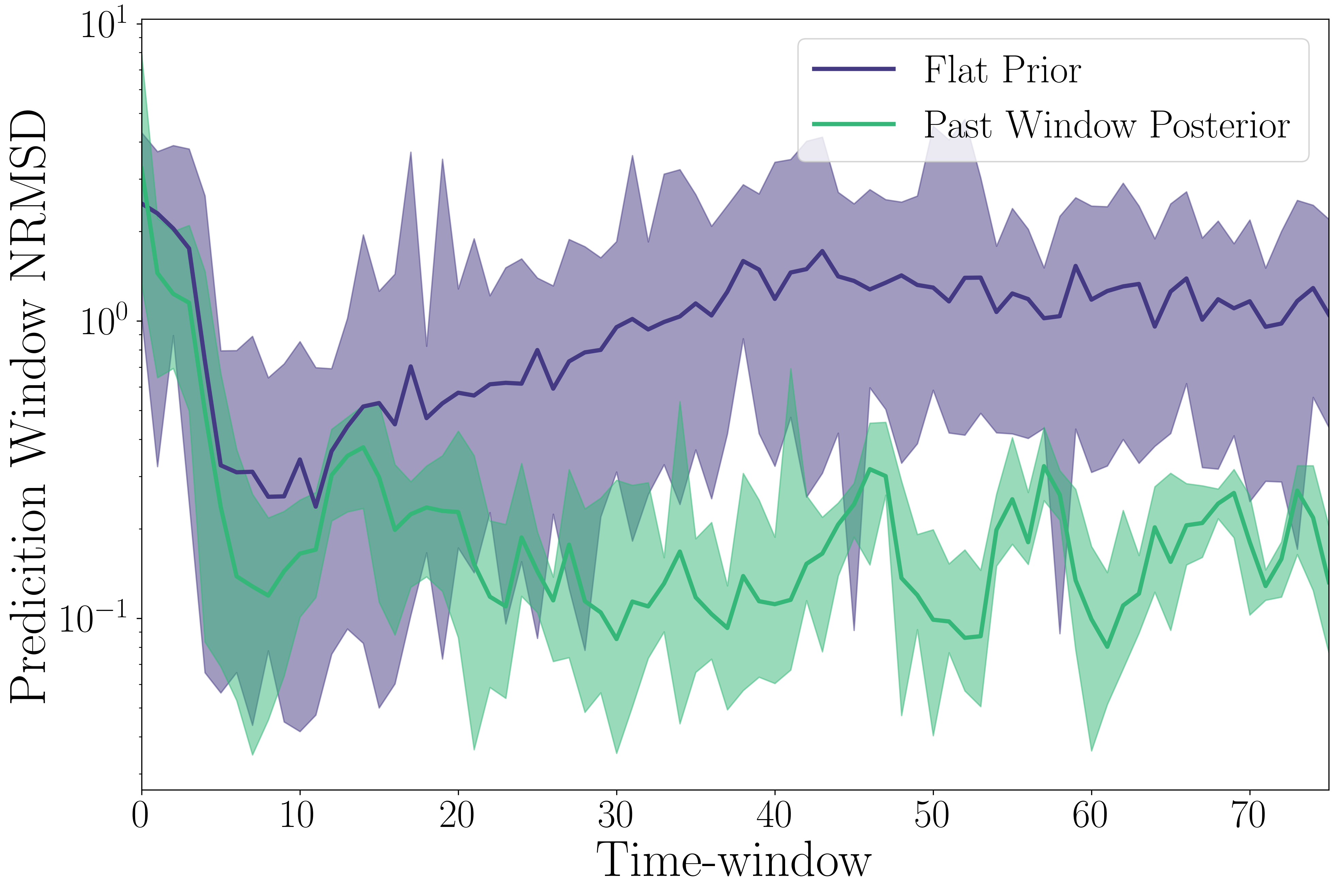}
\caption{Normalized RMSD over the prediction window, for both cases of using flat priors or past window posteriors, considering five executions for each case.}
\label{fig:brazil_30_pred_window_nrmsd}
\end{figure}

We proceed by comparing values of $\varepsilon$ over each time-window considering the quality of both the fit and the prediction. Figures \ref{fig:brazil_30_fit_window_nrmsd} and \ref{fig:brazil_30_pred_window_nrmsd} show the fit and prediction NRMSDs, respectively, for each time-window of the data on Brazil. Using the past window posterior as an informative prior on the current value of the epidemiological parameters leads to an NRMSD approximately two orders of magnitude smaller. During the prediction procedure, the past window posterior approach also shows a smaller $\varepsilon$, this time different by one order of magnitude. Both approaches were fit with 1,000 accepted samples in each posterior of the ABC-SMC algorithm, and the curves are the results of 10 runs of the fitting procedure.

The accumulation of information along the fitting of consecutive time-windows may be analyzed by considering the evolution of $\varepsilon$ through the epidemic data. During the first few time-windows, NRMSD curves in Figures \ref{fig:brazil_30_fit_window_nrmsd} and \ref{fig:brazil_30_pred_window_nrmsd} for the flat prior approach and the past window approach are similar to each other, which indicates that there isn't enough information yet about the parameter's values to be learned by the past window posterior approach. As we fit more time-windows, information is accumulated by the past window approach, leading to smaller NRMSD values.

\begin{figure}[hbtp]
\centering
\includegraphics[width=\linewidth]{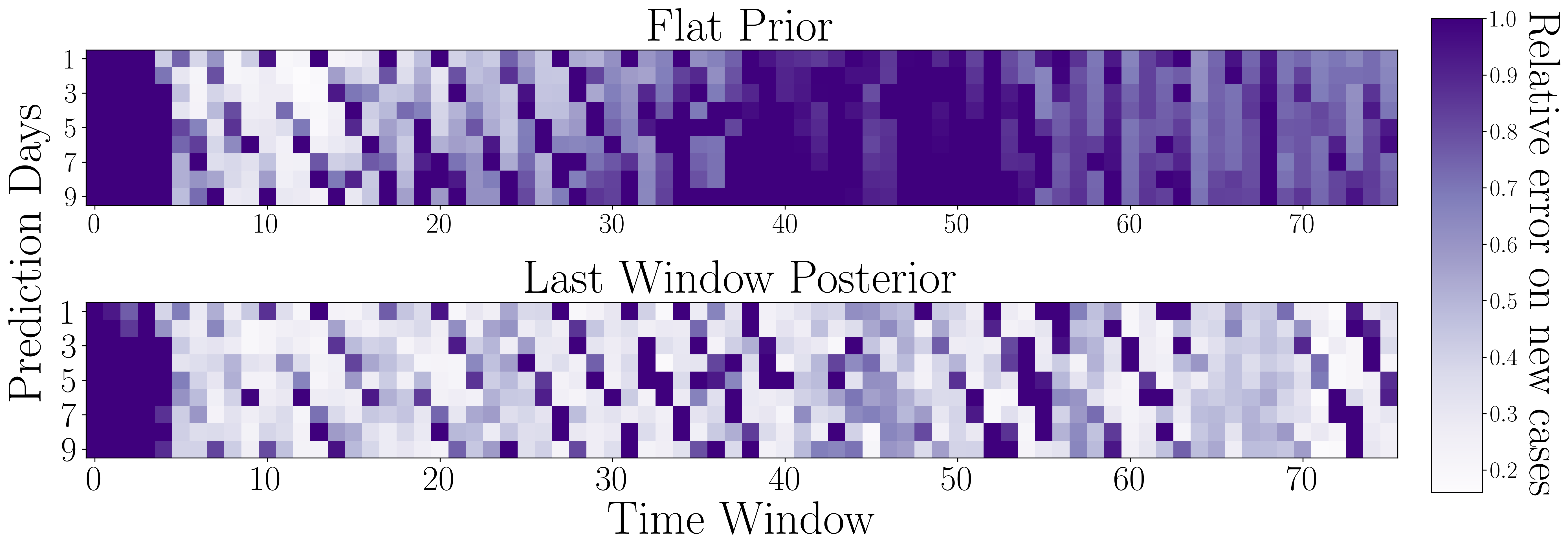}
\caption{Normalized RMSD for every day of the prediction window, for both cases of using flat priors or past window posteriors, considering ten executions for each case.}
\label{fig:brazil_30_heatmap}
\end{figure}

Looking at the prediction error on each day of the prediction window, we get the heat map presented in  Figure \ref{fig:brazil_30_heatmap} comparing the error magnitude for each day of the prediction window in each of the time-windows of the curve. Here, the relative error is calculated as the difference between predicted daily cases and the actual data on it, divided by the data value for normalization. In both cases, the first days show larger errors. However, the past window posterior approach leads to smaller error by day for a longer period. Closer to the 35th time-window, the flat prior approach drastically increases its error through the prediction window (as shown by the purple color). This is further evidence that using the adaptive window method with the past window posterior approach is a more adequate method for generating forecasts for the next few days of the epidemic curve.

\begin{figure*}[hbtp]
\centering
\includegraphics[width=\linewidth]{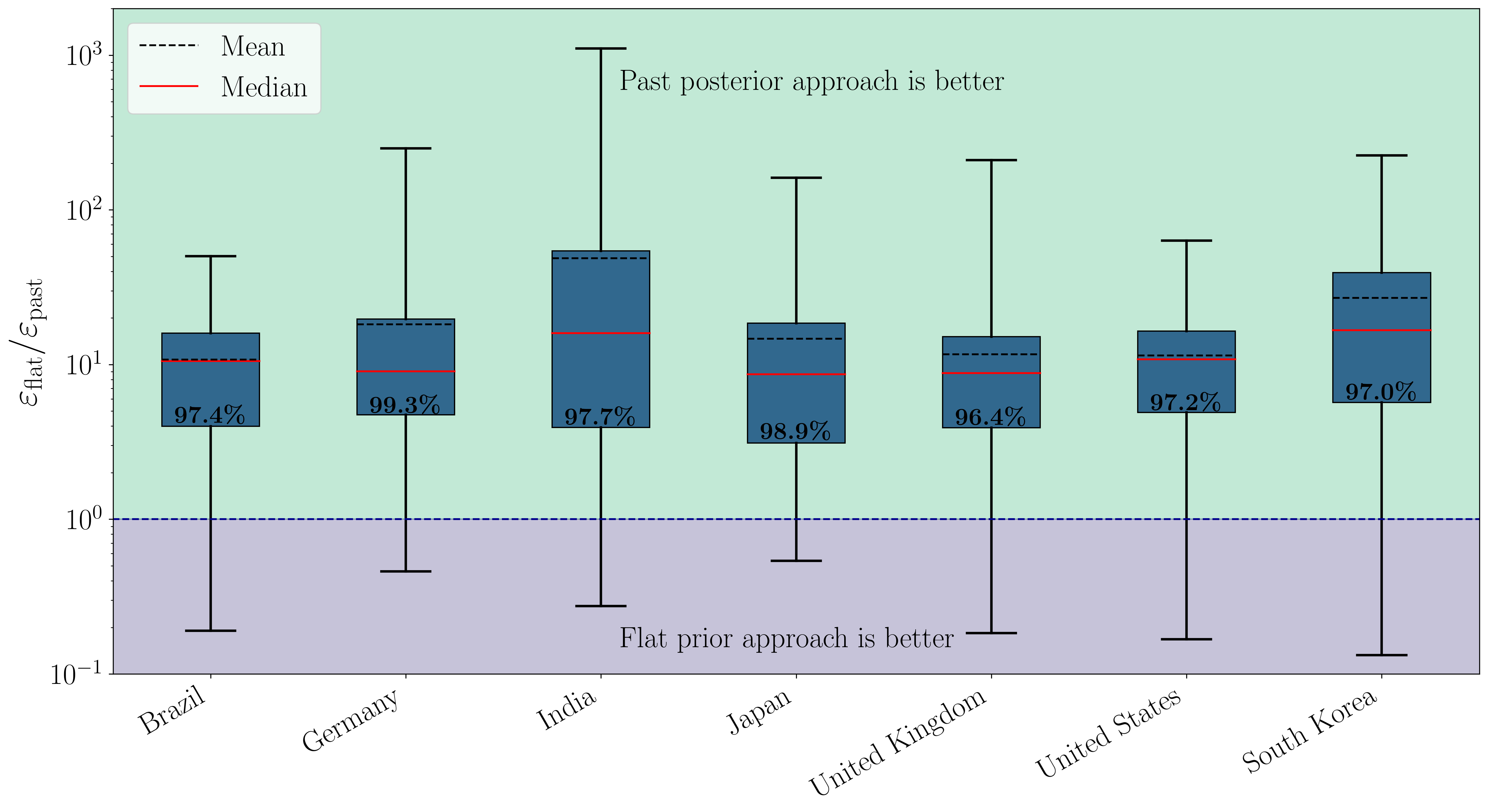}
\caption{Distributions of the ratio of the errors in the fitting period for each country. Points above 1 indicate a preference for the past window approach, whereas points below 1 indicate an advantage for the flat prior approach. In each box-plot, the percentage reflects how many points in the distribution are above 1.}
\label{fig:boxplot}
\end{figure*}

The same analysis presented so far is also done considering data from Germany, India, Japan, South Korea, United States and United Kingdom, and it can be found in section S5 of the Supplementary information. The results remained consistent for other countries, as one can see in Figure \ref{fig:boxplot}, even though the epidemic curves from these countries are quite different from one another, which indicates the robustness of the method. 

Figure \ref{fig:boxplot} shows the box-plots of the distribution of the ratio $\varepsilon_{\mathrm{flat}}/\varepsilon_{\mathrm{past}}$ between fit NRMSDs in each window, obtained via flat prior and past posterior approaches. We considered five different initial window sizes for each country studied, and ten different executions of the inference algorithm. In all countries, over 96\% of the ratio distribution is above 1, indicating a that $\varepsilon_{\mathrm{flat}} > \varepsilon_{\mathrm{past}}$. Therefore, in more than 96\% of the time, the past posterior approach leads to a better model fitting to the data.

We can conclude that separating complex dynamical data in time-windows can allow for its tractability through simple models, and we present a way to do this via approximate Bayesian computation. It is clear that considering data in the past, when choosing the prior distribution for a time-window, leads to better results. In this work, we consider on cases and deaths of COVID-19, but, in principle, this same methodology could be applied in many different scenarios involving dynamical quantities.

\backmatter

\bmhead{Supplementary information}

In the supplementary information text, the reader can find some details of the ABC-SMC implementation used. Also, there is a short study on the practical results of using an adaptive window size, compared to fixed ones, and the consideration of increasing the number of samples used in the flat prior approach, compared to the past window posterior one. Lastly, the results presented in Section \ref{sec:results} are generated also for COVID-19 data from different countries: Germany, Japan, India, South Korea, United States and United Kingdom.

\section*{Declarations}

\bmhead{Acknowledgments}

The authors acknowledge the National Laboratory for Scientific Computing (LNCC/MCTI, Brazil) for providing HPC resources of the SDumont supercomputer, which have contributed to the research results reported within this paper. URL: http://sdumont.lncc.br

\bmhead{Funding}

J. P. Valeriano acknowledges funding by FAPESP, process 2020/14169-0. I.Reis acknowledges funding by FAPESP, process 2021/02027-0. P. H. P. Cintra acknowledges funding by CAPES, process 88887.625345/2021-00. Gustavo Libotte is supported by a postdoctoral fellowship from the Carlos Chagas Filho Foundation for Supporting Research in the State of Rio de Janeiro (FAPERJ), grant number E-26/200.347/2021.

\bmhead{Conflict of interest}

The authors declare that they have no conflict of interest.

\bmhead{Data and Code availability} All the code used to generate the presented results -- together with some examples of the code output for ease of analysis by the interested reader -- is available at the GitHub repository \url{https://github.com/gustavolibotte/LNCC-COVID-19-prediction/tree/TMLearningABC}.



\bibliographystyle{spmpsci}

\bibliography{bibliography}


\end{document}